\documentclass[nameyear]{elsart}
\input psfig.sty

\begin{document}

\newcommand{\lesssim}{\mathrel{
\hbox{\rlap{\hbox{\lower4pt\hbox{$\sim$}}}\hbox{$<$}}}}
\newcommand{\gtrsim}{\mathrel{
\hbox{\rlap{\hbox{\lower4pt\hbox{$\sim$}}}\hbox{$>$}}}}

\newcommand{\bx}{{\vec x}}
\newcommand{\bs}{{\vec s}}
\newcommand{\br}{{\vec r}}
\newcommand{\bq}{{\vec q}}
\newcommand{\bb}{{\vec b}}

\newcommand{\rc}{{$r_c~$}}
\newcommand{\lrc}{{{\rm log}$~r_c~$}}
\newcommand{\muv}{$\mu_V(0)~$}
\newcommand{\vs}{{\em vs.} }
\newcommand{\si}{$\sigma~$}
\newcommand{\lsi}{{{\rm log}$~\sigma~$}}
\newcommand{\s}{{\em S-space }}
\newcommand{\hr}{~{\rm hr}}
\newcommand{\dy}{~{\rm d}}
\newcommand{\cm}{~{\rm cm}}
\newcommand{\m}{~{\rm m}}
\newcommand{\kpc}{~{\rm kpc}}
\newcommand{\Gpc}{~{\rm Gpc}}
\newcommand{\kms}{~{\rm km~s}^{-1}}

\newcommand{\apj}{ApJ}
\newcommand{\apjs}{ApJS}
\newcommand{\araa}{ARA\&A}
\newcommand{\aap}{Astr. \& Ap.}
\newcommand{\mnras}{MNRAS}

\newcommand{\eg}{{\it e.g.~}}
\newcommand{\ie}{{\it i.e.~}}

\begin{frontmatter}

\title{The Fundamental Straight Line of Galactic Globular Clusters
\thanksref{journal}}
\thanks[journal]{accepted by \emph{New Astronomy}}

\author{Michele Bellazzini\thanksref{email}}

\address{Osservatorio Astronomico di Bologna, Bologna, Italy}

\thanks[email]{E-mail: bellazzini@astbo3.bo.astro.it}

\begin{abstract}

A detailed analysis of the Fundamental Plane properties of Globular
Cluster is performed. If a sample of ordinary King-model clusters
is considered, it is found that, in the 
space ({\em S-space}) defined by the parameters 
({\rm log} $r_c$, {\rm log} $\sigma_0$, $\mu_V(0)$),
their configuration is similar to a straight line.  

It is shown that, with rather simple assumptions, {\em a simultaneous 
explanation of all the observed correlations between S-space parameters} can be
provided.  

It is suggested that, at earlier times, 
Globular Clusters populated a line in the three-dimensional
$S-space$, i.e their original dynamical structure was fully determined by a 
single physical parameter. \\

\end{abstract}
\end{frontmatter}

\clearpage

\section{INTRODUCTION}

The study of self-gravitating stellar systems through the analysis of their
distribution into the N-dimensional space of their characteristic parameters
has shown to be a very fruitful tool to learn about the formation and dynamical
evolution of such objects. The method has been succesfully applied to elliptical
galaxies (with the discovery of the so called Fundamental Plane of ellipticals;
\cite{d7}, \cite{dd87}), to spiral galaxies
(\cite{v84}), to galaxy clusters (\cite{sh93}), to globular 
clusters (Djorgovski 1995; hereafter \cite{D95}) and to all these systems at 
once (\cite{bur97}). 

In this scenario, the globular clusters (GCs) deserve a particular place because they
are very simple systems and we also know by direct observations that, at odds
with galaxies, they host a single-age/single-metallicity stellar population and
they have not been subjected to chemical self-enrichment.    

Furthermore, correlations between GCs observables can shed some
light on the processes that led to formation of globulars, a very important
issue by itself (see Meylan \& Heggie 1997, hereafter \cite{mh97}, for a
complete review).

\cite{D95} showed that Galactic globulars are displaced on a plane into the 
three-dimensional space
defined by the {\rm log}arithm of their core radii (\rc, in pc), V-band central 
surface 
brightness (\muv, in mag/arcsec$^2$) and {\rm log}arithm of the velocity dispersion 
(\si in km/s) 
- hereafter {\em S-space} - i.e. they constitute a bidimensional manifold in
{\em S-space}. The corresponding scaling law indicates that GCs
cores are virialized systems with constant mass to light ratio ($M/L$).
In these hypotheses, the generic condition sufficient to obtain the observed
Fundamental Plane of Globular Cluster (GCFP) is that the scatter induced in 
the GCFP by non-homo{\rm log}y between globulars is smaller than that produced by
the dispersion in M/L around a mean value, and by observational errors. 
This is certainly the most interesting among the three conditions (i.e.,
virial equilibrium, constancy of M/L and homo{\rm log}y; see \cite{ciotti})
but, for the aim of the present analysis, it can be considered as an
observational fact.
         
\cite{D95} showed also that if a space defined by all the photometric,
structural and dynamical parameters is considered - hereafter SE-space -, the
dimensionality of the manifold is still not more than 3: that means that many
of the involved parameters correlates between each other (see also Djorgovsky
\& Meylan 1994 (\cite{DM94}), \cite{mh97}). Many of these correlations are not trivial and
the origin of most of them is still unclear \cite{DM94}. 
However Bellazzini et al. (1996;
\cite{BVFF}) and Vesperini (1997; \cite{v97}) demonstrated that, for instance, the
correlation between integrated cluster magnitude ($M_V$) and the {\rm log}arithm of
the central luminosity density (${\rm log} ~\rho_0$) is very likely to have been
settled at the time of GC formation, and that dynamical evolution, either
intrinsic or due to the interaction with the host galaxy, is much more
efficient in {\em damaging} any existing correlation between structural
parameters than in settling one such correlation by means of evolutionary or
selective disruption effects. These latter studies showed that it is possible
to recover informations on the initial dynamical/structural conditions of a
GCs system from present-day correlations between {\em SE-space} parameters. 

In this short note I present a number of hints suggesting that Galactic GCs
were formerly displaced on a Straight Line in the {\em S-space},
i.e. they formed a one-dimensional manifold in such space. 
Furthermore I present a simple and general explanation for all
the observed mono-variate correlations between \s parameters, comprehensive of
a new, more satisfactory, interpretation of the correlation 
between cental surface brightness and velocity dispersion. 

\section{The S-space manifold of King Model globular clusters}

\cite{D95} exploited the existence of the Fundamental Plane of Galactic Globular
Clusters (GCFP) in the \s by {\it a)} demonstrating that the bi-variate GCFP
correlations were significantly better than any of the mono-variate ones
intercurring between each couple of the \s parameters and, {\it b)} showing,
with the application of Principal Component Analysis methods\footnote{Principal
Component Analysis (PCA) is a multivariate analysis statistical technique apted
to find the true dimensionality of a given dataset in a N-dimensional space of
parameters. PCA finds the eigenvectors (principal components) that maximize 
the variance of
data-points in a given space: often the sum of the variances accounted by 
the first M eigenvectors (with $M < N$) equals the whole cosmic variance 
expected in the dataset (the remaining variance being accounted by the
``observational noise'' present in the data). 
That means that data-points lie on a manifold of 
dimension M in the considered space, i.e. any of the N parameters correlates 
with a linear combination of M others, i.e. there are N-M couples of parameters
that present strong correlations, thus are nearly equivalent for the description
of the dataset properties. A brief and clear introduction to PCA can be found in
\cite{D95}. For deeper insights see \cite{muhec}}, that ``... the first two
principal component of the data ellipsoid account for the $98.4\%$ of the total
sample variance ...''. In other words, the amount of sample variance that could
be accounted by any further dimension (for example, a third principal
component) is significantly less ($1.6 \%$) than the variance that is expected
to be introduced into the sample by mere observational errors (which has to be
at least $\sim 10 \%$, given the current uncertainties in the estimate of the
involved observables; see Djorgovski 1993a, hereafter \cite{D93}; see also
sec. 3, below).     

However, the sample analyzed by \cite{D95} include both normal clusters, whose
surface brightness profiles are well fitted by ordinary King models \cite{k66},
and
also clusters presenting a power law cusp at their centers, and consequently 
classified as probably having passed the core collapse phase of their dynamical 
evolution (i.e. classified ``c'' or ``c?'' by \cite{D93}; see also \cite{DM94}, \cite{mh97} and 
references therein). Most of the
globular cluster structural parameters refers to King models and so are somehow
ill-defined when applied to Post Core Collapse Clusters (PCC); for instance 
\cite{D93}
is forced to assign a conventional value of cluster concentration
($C={\rm log}{r_t\over{r_c}}=2.5$, where $r_t$ is the cluster tidal radius 
\cite{k66} because this parameter is defined {\em by} King models. 
Concerning the
\s parameters, at least one of them is uncertainly defined for PCCs, i.e. core
radius (see \cite{BVFF} for a wider discussion). 
So, inclusion of PCCs into the scatter plots involving
\s parameters may be somehow risky or misleading. Furthermore, even if listed
core parameters for PCCs were reliable, these clusters have passed the
most evolved phase of their evolution and most of the records of initial
conditions has probably been erased during the collapse.   

The
three mono-variate correlations between the \s parameters and two GCFP
bi-variate correlations, also shown by \cite{D95}, are reported in fig. 1. 
Both the data sources (\cite{D93}, \cite{PM93}, 
Trager, Djorgovski \& King 1993, hereafter \cite{TDK93}) and the scales of the 
plot are 
exactely the same of \cite{D95}, but in the present diagrams PCCs are plotted as
filled squares while open squares correspond to ordinary King Model Clusters
(KMC). The only difference with \cite{D95} is that, in the \lrc {\em vs.} \muv
plot, I
included in the sample all the clusters that have a measure of both of these
parameters (also when \si entry is missing) in the \cite{D93} compilation: this choice, 
that has been followed all over
this work, cannot affect the presented result. The use of a larger sample, at
least when this specifical plane is considered, can only increase the
statistical significance of the results themselves. 

From the inspection of fig. 1, it is readily evident that much of the 
dispersion in the mono-variate
plots is due to PCCs and that, if only KMCs are considered, the mono-variate
correlations are strong and significant and, at least for two of them [(\muv \vs 
\lsi) and (\lrc \vs \muv)], their quality is comparable with that
of the corresponding bi-variate ones. Since both \lsi and \lrc are strongly
dependent on \muv they {\em cannot} be truly independent, so the relatively
high dispersion of the data-points into the (\lrc \vs \lsi) plane can also
derive from a perverse composition of the observational errors (but see sec. 3.2
for a more satisfactory explanation).

If each parameter of a given N-dimensional space correlates with any of the N-1
others it means that the sample is displaced along a straight line into this
space. Though PCA indicates that the statistical dimension of the manifold of 
KMCs is probably still 2 , the situation is indeed suggesting the existence of a 
Fundamental Straight Line (FSL) of KMCs. 
The first eigenvector accounts for $84.5 \%$ of the 
KMC sample variance while, if PCCs are included in the sample, the variance
accounted by the first eigenvector is only $76.0 \%$. In the upper right panel
of fig.1 (\lrc \vs \lsi) there is a point in clear disagreement with the
trend shown by the others KMCs, in the
corner corresponding to high \lrc and high \lsi: this correspond to NGC 5139
($\omega$ Cen) which is known to be a very anomalous cluster from a dynamical
point of view (\cite{m87}, \cite{WS87}) and also peculiar under many 
other aspects (\cite{mh97}, and references therein); on these bases doubt have been
casted on its very classification as a globular (\cite{m96}). If NGC 5139 is
excluded from the KMC sample the amount of variance accounted by the first 
eigenvectors grows to $87.5 \%$, i.e. the expected amount of cosmic variance in
the dataset, given the current uncertainties in the observables.

Finally if, according to the approach introduced by \cite{BVFF} and \cite{v97}, we 
consider
the two subsample composed of KMCs that find themselves {\em inside} the Solar
Circle [$R_{GC}\le 8 Kpc$ -- {\em Inner Clusters} (IC)], and outside 
this region [$R_{GC} > 8 Kpc$ -- {\em Outer Clusters} (OC)], we find that
the dimensionality of the OC sample is unambiguously {\em one}, the first
eigenvector accounting for $91 \%$ of the sample variance, while IC
data-points are less clustered along their first eigenvector, which
still represent a $74 \%$ of the IC sample variance. So there is a marginal
evidence that the clusters that are expected to be less affected by
dynamical evolution (i.e. OC ones, see \cite{mh97}, \cite{BVFF} and \cite{v97}) define the best
mono-variate correlations in \s. 

In the following analysis PCCs will be permanently 
excluded from the sample, and the presented results refer only to KMCs.

\begin{figure}[htbp]
\centerline{\psfig{file=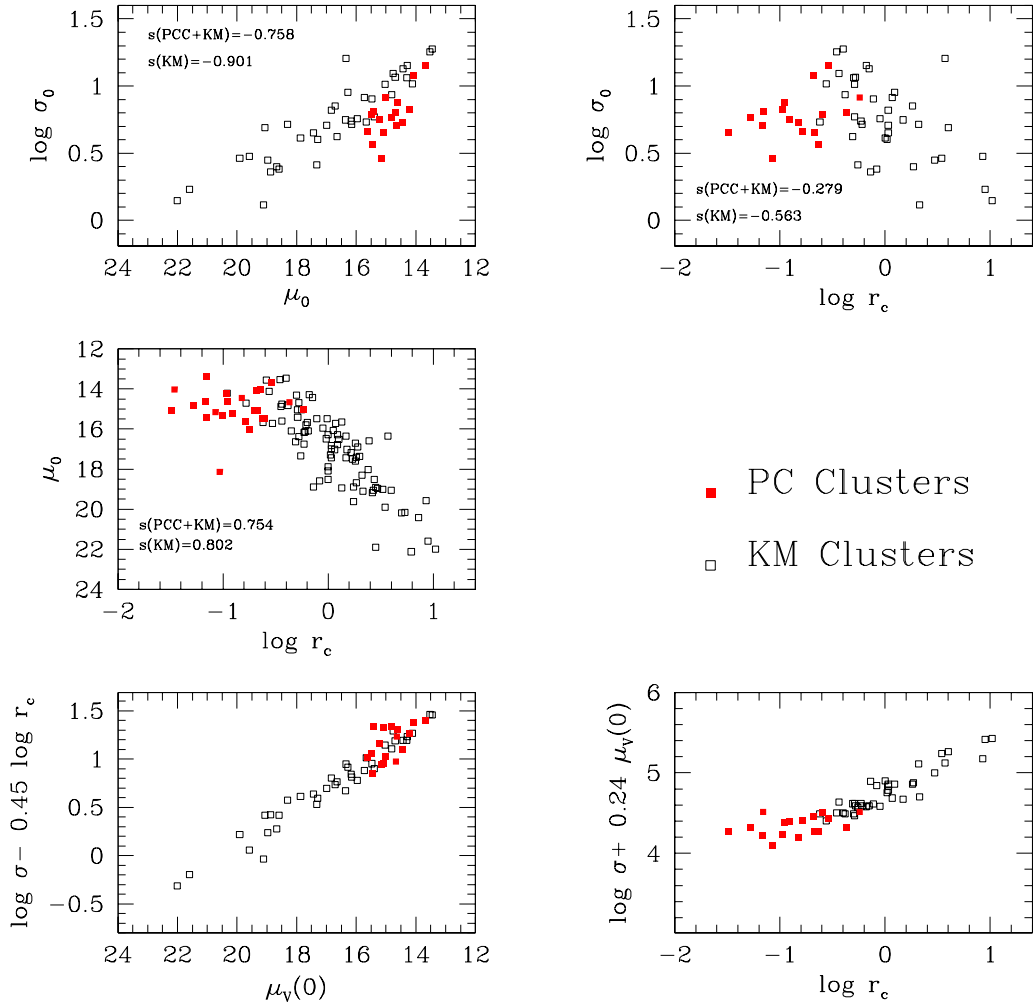,width=5.9in}}
\caption{Mono-variate correlations between \s parameters (top two panels and
middle panel). Spearman rank correlation coefficient (see \cite{BVFF}) is reported in
each plot for the whole sample (KMC+PCC) and for the KMC sample only. It is
evident that the correlations in the latter sample are much stronger than in
the KMC+PCC one. The two lower panels display, for comparison, the GCFP 
correlations found by \cite{D95}. In each of the five plots filled squares represent 
PCC while open squares represents KMC. 
}
\label{fig:pk}
\end{figure}

\subsection{Relations between manifolds in S-space}

Simple geometrical considerations show that the eventual FSL defined by KMC,
{\em belongs}, whithin the errors, to the GCFP. Roughly speaking,
the two correlations (\muv \vs \lsi) and (\lrc \vs \muv) are nearly-edge-on
view of the GCFP, while the (\lrc \vs \lsi) correlation is a nearly-face-on
view of the same Plane. The main result of the present section is that the
surface of the GCFP is not uniformly populated by KMC (the large majority
of galactic globulars): the clusters are confined in a rather dispersed but
well defined band which represents the correlation between {\rm log} of the core 
radii and {\rm log} of velocity dipersions.  

A similar result, regarding the half-light parameters (see \cite{DM94} for
definitions), has been very recently suggested by Burstein et al. (1997). 
They claim that the GC
half-light parameters manifold has been reduced to a straight line by selective
disruption mechanisms(see \cite{go97}, and \cite{mw97}), which depleted those 
clusters whose radii lie out of a narrow ``permitted'' range. 
I will show below that this explanation is unlikely to be the right one in the \s 
considered here.

In the present work I did not
include half-light parameter in the analysis for a number of reasons, the 
main one being that I am mainly interested in discriminating between 
"evolutionary" \vs "initial condition" origin for the observed correlations.  
The bulk of the GC evolution takes place in their core - they are
the gravitational engines driving the life of the whole system (see 
\cite{spitz}) - so core parameters are expected to be more sensitive indexes of the
dynamical status of the clusters themselves.

\section{At the root of S-space correlations}

In the following sections I attempt a deep analysis of all the
detected \s correlations, either bi-variate (FP) or mono-variate (FSL),
with particular attention to the limitations in our understanding of the
underlying physics that can be imposed by the parametrization currently
used, i.e. the observables. Finally I show that a very simple structural
condition of KMCs provides a simultaneous explanation of all the FSL
correlations.

First of all, I want to draw attention to the uncertainties in the
observables. The internal errors of the data in the KM sample are
between $0.2$ and $0.4 dex$ for \muv and $0.07$ to $0.14 dex$ for \lrc, 
as specified
by \cite{TDK93}whose catalogue is at the origin of the \cite{D93}
compilation. \lsi is the most critical quantity: while the typical uncertainty
is of the order $0.1$ -- $0.2 dex$, relative errors equal or higher than $ 100
\%$ are not rare, so \lsi errors can amount to $0.4 dex$ or more.

Even so, statistical uncertainties are not necessarily the major source of
concern for the present purposes. In fact:

\begin{itemize}

\item \si measures have been obtained with very different methods: either
from individual radial velocities of cluster members or from the Doppler
broadening of an integrated light spectrum. The data have very
heterogeneous sources and ``...no attempt (was made) to remove zero-point
differences between the different sets of radial velocities...'' (\cite{PM93}). 
So some additional systematic uncertainty is indeed present in
the data.

\item Regardless of the method adopted for measuring \si, the
cluster distance, the degree of crowding and the instrumental set up
of each observation impose some constraints upon the region of the
cluster in which the measure is performed. Individual stellar velocity 
measures are often possible only outside of the badly crowded cores
of many clusters. Integrated light spectra are not affected by this kind
of problem, but it is readily evident that, for a fixed instrumental
set-up, the contribution of light from outer cluster regions will increase
with increasing cluster distance and will also depend on the cluster concentration
(C). So, it is not at all
guaranteed that each of the measures in the dataset refers to the same
quantity and, in particular, to the quantity of interest, i.e. ``average
velocity dispersion in the core'' (\cite{D95}).

\item Heterogeneity of sources can significantly affect also the
internal consistency of \muv and \rc values. These parameters are derived
from radial brightness profiles obtained with very different methods,
ranging from concentric aperture photometry to direct star counts. In some
cases more than one method was used to track the profile of the same cluster
in different radial regions. Though the datasets of \cite{TDK93}
and \cite{D93} represent the result of an enormous effort and it is certainly the
best result presently achievable in terms of homogeneity and completeness,
some undetermined internal inconsistency should still be present, as 
clearly stated by the same authors.

\end{itemize}

Given the above considerations, it is somehow {\em surprising} that the correlations
shown in fig. 1 are nevertheless observed. 
It should also be noted that the amount of variance introduced 
in the dataset by global ``observational noise'' can be significantly greater
than estimated taking into account only formal uncertainties of each
observable. 

\subsection{M/L ratio from GCFP correlations}

Let me first consider GCFP correlations. The basic conditions necessary to the
existence of the observed GCFP correlations are {\em a)} the obvious one
that GC cores obey the Virial Theorem and {\em b)} 
that M/L ratio\footnote{Since all luminosity observables refers to V filter 
measures, the M/L ratio must be intended as $M/L_V$ - in solar units - all 
over this paper, the subscript V having been dropped for brevity} be nearly
constant (see \cite{D95}, \cite{ciotti}). Using Virial Theorem, 
the mass of the cluster cores ($M_c$) can be estimated, according to the 
prescription adopted by \cite{bur97}, with the formula:   

\begin{equation}
M_c= r_c \sigma^2 C_a^2/G~~~~~ 
\label{eq1}
\end{equation}

where $C_a$ is a constant (here assumed to be equal to $\sqrt{2}$, according to
\cite{bur97}) and G is the universal
Gravitational constant. Converting to suitable units and in {\rm log}arithmic
form:

\begin{equation}
{\rm log} M_c = 2 {\rm log} \sigma + {\rm log} r_c + 2.67~~~~~ 
\label{eq2}
\end{equation}

with $M_c$ in solar masses, $\sigma$ in Km/s$^{-1}$ and \rc in pc.

The same quantity can be obtained from \muv and \rc, once assumed a fixed M/L
ratio, from $L_c = C_b \pi I_0 r_c^2 $ (\cite{DM94}), 
where ${\rm log} I_0 = 0.4(26.362 - \mu_V(0))$ and $C_b$ is a constant (here assumed
to be equal to 1; see, for instance \cite{DM94}), through the equation:

\begin{equation}
M_c = \pi C_b I_0 r_c^2 (M/L)~~~~~ 
\label{eq3}
\end{equation}

that, transformed as above gives:

\begin{equation}
{\rm log} M_c = -0.4 \mu_V(0) + 2 {\rm log} r_c + {\rm log} (M/L) + 11.04 ~~~~~ .
\label{eq4}
\end{equation}

Eliminating $M_c$ between (\ref{eq1}) and (\ref{eq3}), (that is imposing that the
``dynamical core mass'' be equal to the ``luminous core mass'') it is found:

\begin{equation}
{\rm log} r_c = 2 {\rm log} \sigma + 0.4 \mu_V(0) - {\rm log} (M/L) - 8.37 ~~~~~ 
\label{eq5}
\end{equation}

or, in another form:

\begin{equation}
\mu_V(0) = -5 {\rm log} \sigma + 2.5 {\rm log} r_c + 2.5 {\rm log} (M/L) + 20.92 ~~~~~ .
\label{eq6}
\end{equation}

The coefficients of equations (\ref{eq5}) and (\ref{eq6}), as expected, are in 
excellent agreement with those of the fits to observed GCFP correlations (\cite{D95}). 
I have obtained the GCFP relations in a parametric form, with {\rm log} (M/L) as
a parameter, from the simple equality between luminous mass and dynamical mass. 
However, there is one more implicit
assumption, i.e. structural and kinematical homology of GCs (\cite{ciotti};
see also \cite{D95} for a simple explanation referred to GCs). Some effect of
non-homo{\rm log}y should necessarily be present since , for instance, density
profiles of globulars are observed to differ from one another by more than a
simple scale factor (\cite{D95}, see also \cite{spitz}, p. 16), 
so slight differences in the $C_b$ constant are expected. This condition alone
has to produce also cluster-to-cluster differences in the $C_a$ value 
(\cite{ciotti}). 
 The very existence of the GCFP
demonstrates that non-homology is not a major concern in our case, nevertheless
it surely contributes to the dispersion of data-points about the Fundamental
Plane (and about FSL too).

\begin{figure}[htbp]
\centerline{\psfig{file=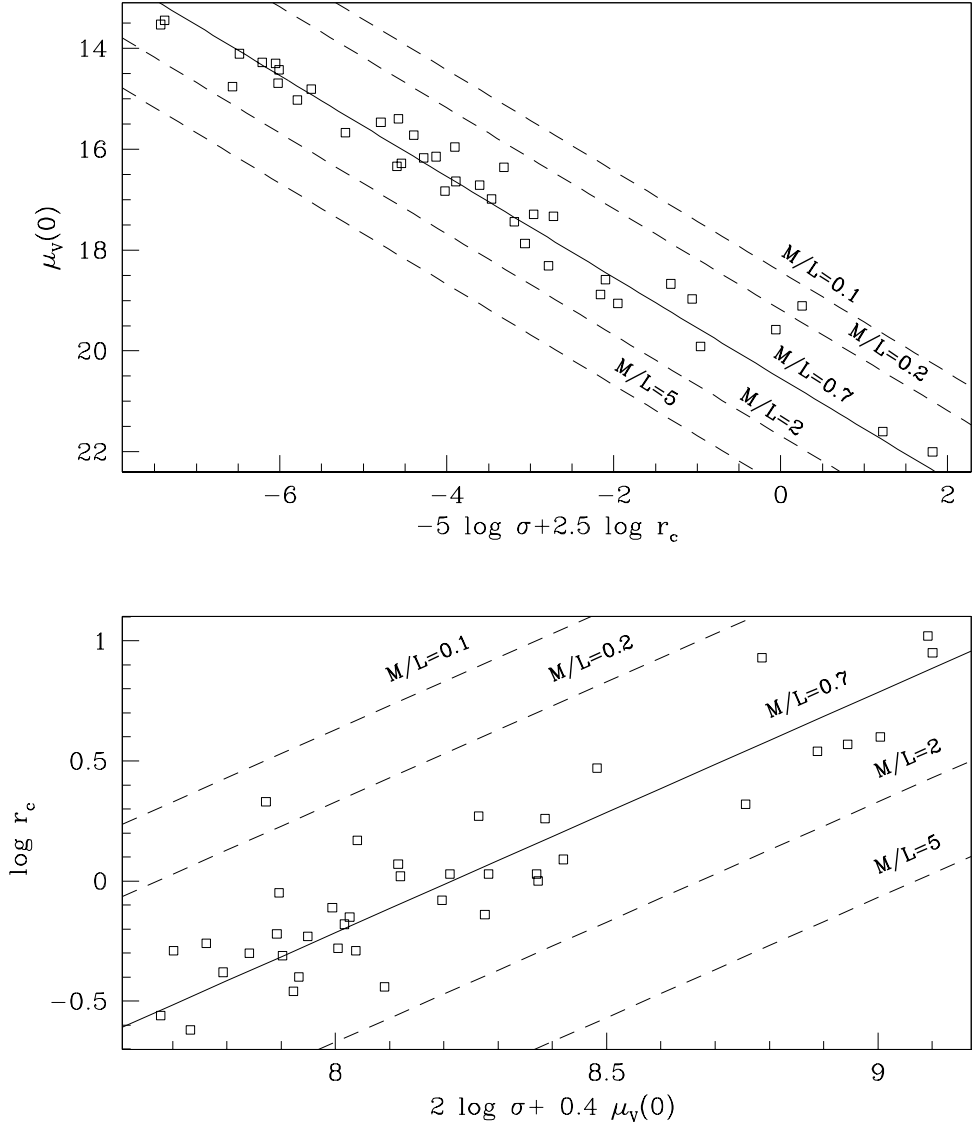,width=6.2in}}
\caption{ Edge on views of the GCFP. The superimposed
lines are {\rm log} (M/L) = constant realizations of equations 
(\ref{eq5}, lower panel) and (\ref{eq6}, upper panel). Note that
the slope of the lines excellently match the apparent slope of the distribution
of data-points. While the actual value of M/L as read from the line that best
represents the data is probably not correct (but still consistent with current
estimates) the plots give a realistic estimate of the dispersion in M/L between
GCs as measured from present-day data. The outlier point above the $M/L = 0.2$
line corresponds to NGC 6366, a cluster affected by a relatively high
foreground extintion and with a quite uncertain measure of the velocity
dispersion (see \cite{PM93}, who obtain for this cluster $M/L_0 = 0.4$).  
}
\label{fig:pk2}
\end{figure}

Two different edge-on views of the GCFP (the same presented by \cite{D95}) are 
shown in fig. 2, with the lines corresponding to equations (\ref{eq5})
and (\ref{eq6}), for different values of the M/L parameter, superimposed to the plot. 
The line which provide the apparent best fit
to the distribution of the data-points corresponds to $M/L = 0.7$ 
in both the considered planes . 
Though in reasonable accord with current
(model dependent) estimates ($<(M/L)_0> = 1.7 \pm 0.9$, Pryor \& Meylan 1993; see
also \cite{cw} for theoretical estimates) this estimate is 
probably not correct, the zero-point of equations (\ref{eq5}) and (\ref{eq6}) being affected 
by many relevant uncertainties. For instance, the assumed $C_a$ value can be
significantly different from its {\em true} average value; an underestimation 
of $C_a$ by a factor $\sqrt{2}$ lead to an underestimation of the M/L ratio by 
a factor of $ \sim 2$.  

What, on the other hand, is much more robust is the estimate of the 
{\em spread} in M/L: a factor $\sim 10$ can be viewed as a strong upper limit. 
Considering
the many sources of statistical and systematic error that can contribute to the
observed dispersion, one is induced to argue that the real spread is lower
(perhaps much lower), and
indeed the 1-$\sigma$ dispersion is a factor $\sim 2.2$. However the thickness
of the GCFP is comparable to that of the Fundamental Plane of Elliptical
Galaxies (FPEG). At any given mass the 1-$\sigma$ spread of data-points around 
the GCFP, as measured by the $k_3$ parameter (which measures {\rm log} M/L; see
\cite{bur97} and references therein), is $\simeq 0.1 ~dex$ (see \cite{ciotti},
and references therein for comparisons with FPEG). 

Summarizing the results of the above analysis, I have obtained:

\begin{enumerate}

\item an estimate of the spread in M/L between clusters;

\item an estimate of the mean M/L as a parameter of equations (\ref{eq5}) 
and (\ref{eq6}). This
will allow me to derive core masses either from equation (\ref{eq2}) or (\ref{eq4}) 
in a fully self-consistent way, a very useful condition for the arguments developed 
in the following sections. 

\end{enumerate}

\subsection{A unique interpretation for FSL correlations}

A natural interpretation for the correlation between \muv and \lrc is that the
core luminosity (and core mass, once $M/L=const.$ is assumed) is nearly
constant. Such an explanation has been
proposed by Djorgovski (1991, 1993b), and it is widely discussed in \cite{DM94}. 

Equations (\ref{eq2}) and (\ref{eq4}) can be used independently to express this condition.
Self-consistency between the two can be imposed assuming $M/L = 0.7$, as shown
above. Solving equation (\ref{eq2}) with respect to \lsi and equation (\ref{eq4}) with 
respect to \muv the following relations are found:  

\begin{equation}
{\rm log} \sigma = -0.5 {\rm log} r_c + 0.5 {\rm log} M_c - 1.33 ~~~~~ 
\label{eq7}
\end{equation}

and

\begin{equation}
\mu_V(0) = 5 {\rm log} r_c + 2.5 {\rm log} M/L -2.5 {\rm log} M_c + 27.5~;~ with~M/L=0.7.
\label{eq8}
\end{equation}

Combining the two equations, a relation between \muv and \lsi can also be
found:

\begin{equation}
{\rm log} \sigma = -0.10 \mu_V(0) + 0.25 {\rm log} M/L + 0.25 {\rm log} M_c + 1.41~;~with~M/L
=0.7 .
\label{eq9}
\end{equation}

Just using the hypothesis that $M_c = const.$ (and M/L fixed), I have obtained 
three mono-variate parametric relations between the \s observables, 
with {\rm log} $M_c$ as a parameter.

In fig. 3 the lines corresponding to equations (\ref{eq8}) (left panel), 
(\ref{eq7}) (central
panel), and (\ref{eq9}) (right panel), with different values of the 
{\rm log} $M_c$ parameter, are superimposed to the respective FSL 
correlations. It is immediately evident that {\bf the sole} (quite broad)
{\bf condition that $M_c$ be constant within two orders of magnitude is 
sufficient to impose all the observed correlations}.

\begin{figure}[htbp]
\centerline{\psfig{file=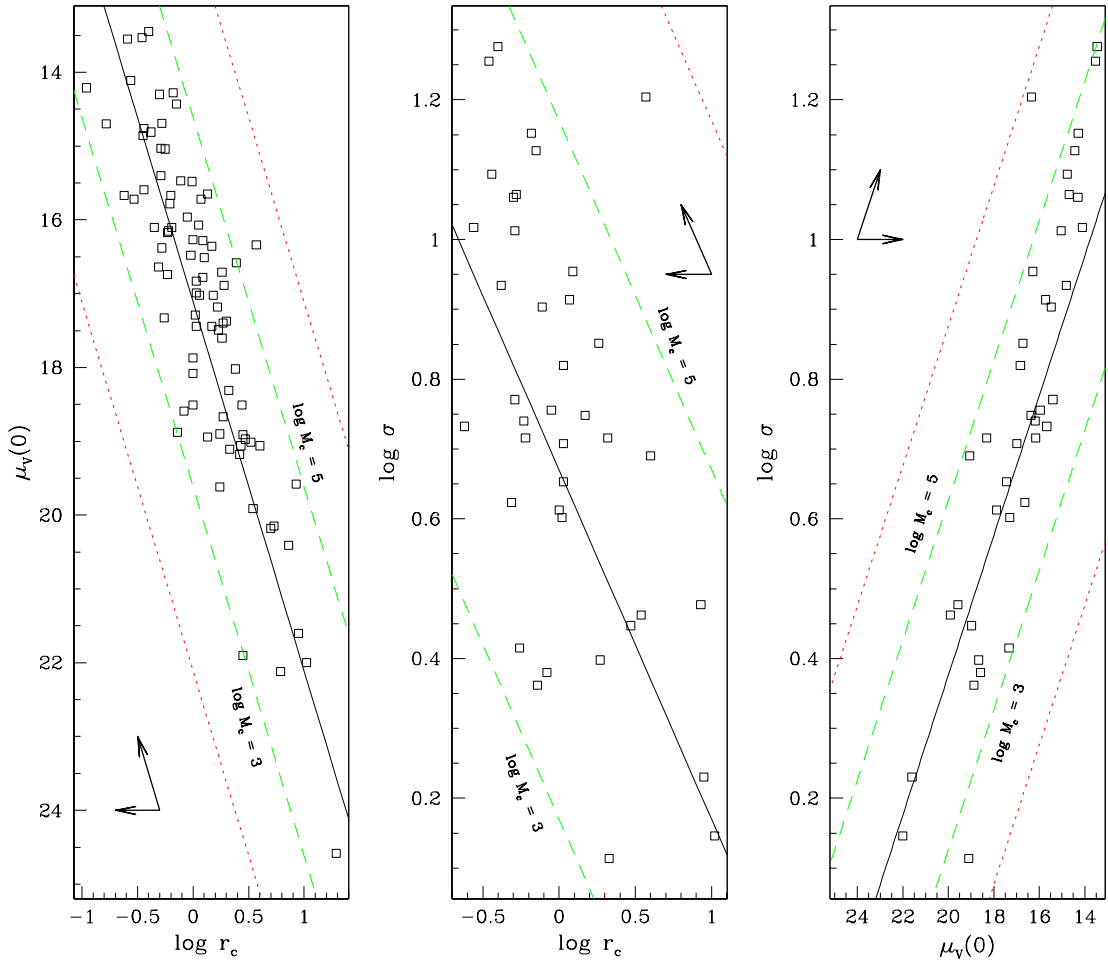,width=5.9in}}
\caption{
The three mono-variate correlations between the \s parameters. The superposed
lines represents respectively equations (\ref{eq8}) (left panel), (\ref{eq7}) (central panel),
and (\ref{eq9}) (right panel), for different values of the parameter {\rm log} $M_c$ and M/L
fixed to 0.7. The outlier point [above the ({\rm log} $M_c = 5$) line] is 
NGC 5139.
The central continuous line corresponds to {\rm log} $M_c = 4$, while the most
external dotted lines correspond to {\rm log} $M_c = 2$ and {\rm log} $M_c = 6$ and roughly
confine the region of the planes that is {\em permitted} to KM clusters. The
line {\rm log} $M_c = 2$ is out of scale in the central panel. 
It is clearly evident that the simple and broad condition that $M_c$ be constant
{\em within two orders of magnitude} is sufficient to impose all the observed
correlations. Note that the relations presented in the central panel is
completely independent from the M/L ratio.
The couple of arrows constraints (within the acute angle between them, in each 
of the three plane), the possible directions of the ``evolution vectors'', i.e.
the directions along which the points can move into each plane due to ordinary
evolutionary processes (see text). 
}
\label{fig:pk3}
\end{figure}

While the actual values of {\rm log} $M_c$ may suffer from the same uncertainties
discussed above for M/L, they can be considered at least indicative. The dotted
lines in each panel of fig. 3 represent the conditions {\rm log} $M_c = 2$ and 
{\rm log} $M_c = 6$ , i.e. they approximately confine the region in which globulars
are allowed {\em to exist}, as King Model clusters. The upper limit 
roughly correspond
to the very strong condition that core mass of a cluster cannot be higher than 
its total mass, the lower limit is connected whith the decreasing of core mass
occurring during the evolution of globulars toward core collapse (Spitzer 1987,
\cite{cw}, \cite{mh97}). It can be assumed that below a certain
threshold of $M_c$, here arbitrarily assumed to be $10^2 ~M_{\odot}$, the core
collapse of the system has already happened.

It is interesting to imagine a hypothetical researcher
measuring \s parameters for a system of - say - 10000 KM globular clusters with
nearly constant M/L ratio and
with core masses ranging from $10^2 ~M_{\odot}$ to $10^6 ~M_{\odot}$. 
When plotted into the  (\muv -- \lsi) and (\lrc -- \muv) planes, all of the
measured points would be confined between the quoted dotted lines and our
imaginary astronomer would be forced to conclude that a strong (though
significantly dispersed) correlation is present in both planes; the 
interpretation of the (\muv -- \lsi) correlation would represent a considerable
challenge. It would be not be so readily evident that he is observing a 
{\em mere condition of existence!}

Such a tricky situation is due to the following main reasons:

\begin{enumerate}

\item The thickness of the (\muv -- \lsi) and (\lrc -- \muv) correlation is
constrained , in the studied case, by the very existence of the GCFP, i.e. by
the condition $M/L = const.$ Whithout this additional constraint the relative
constancy of $M_c$ would not be sufficient to settle the observed correlations.

\item The definition of photometric magnitude artificially expand the scale of
the \muv axes by a factor 2.5 that is (obviously) missing in the other \s
parameters.

\end{enumerate}

These are also the reasons why the condition $3 < {\rm log} M_c < 5$ encompasses a much
broader region into the plane (\lrc -- \lsi), which is unaffected by the above
constraints.

\subsubsection{Physical constraints into the S-space}

Now I can shortly turn again to the relations between \s manifolds, considering
the FSL as a true manifold. Though its statistical significance is not
rigorously defined, the fact that data-points are distributed along a {\em fat} 
straight line is undoubtely observed.  This means that {\em two} independent 
physical constraints are driving the characteristics of the dataset:

\begin{enumerate}

\item The most powerful constraint is the constancy of the M/L ratio between
globulars within a factor of few. This condition alone (but see the discussion
in sec. 3.1) define the GCFP which is a much more statistically significant
structure with respect to FSL.

\item Whithin the GCFP, globulars are constrained to lie within a strip
corresponding to the constancy of $M_c$ within a factor of $\sim < 100$. This
condition is significantly weaker than that regarding M/L, however each of the
mono-variate correlations of fig. 3 would be much more dispersed than what 
presently observed, if a wider range of $M_c$ were allowed. 

\end{enumerate}

\section{A deeper insight into the FSL correlations}

As said, the position $M_c \simeq const.$ is at the base of the standard
interpretation of the (\lrc -- \muv) correlation (\cite{DM94} and references therein).
On the other hand {\em a)} I am not aware of any discussion of the correlation
between \lrc and \lsi (but this is the less significant one and regards only
KMCs) and {\em b)} the proposed interpretations for the (\muv -- \lsi) are 
different from that presented here (\cite{DM94}). In this section I will shortly
comment on point {\em b}  and add few considerations regarding the effects of
the dynamical evolution of globular clusters on the \s correlations.

\subsection{\muv vs. \lsi}

\cite{DM94} attempt to interpret the correlation between total luminosity (L) 
and velocity
dispersion claiming that in a primordial phase of GC evolution, dominated by
adiabatic mass loss due to stellar evolution, the quantities $MR$ and $R\sigma$
were adiabatic invariants ( where M, R and $\sigma$ are the characteristic
mass, radius and velocity dispersion af the clusters - see also \cite{D91}, 
and \cite{D93b}). 
These conditions should have settled the scaling law $L \propto \sigma$: \cite{DM94}
invoke differential effects in subsequent evolution as responsible of changing
the slope of this relation to the presently observed value, i.e. $\sigma
\propto L^{\sim 0.6}$ (see Vesperini 1994 for a critical discussion).
Turning to surface brightness, assuming the constancy of 
$MR$ and $R\sigma$ (plus $M/L =const.$), the scale relation 
$\sigma \propto I_0^{0.33}$ is derived, similar to the observed one 
($\sigma \propto I_0^{0.5 \pm 0.1}$ including PCC in the sample; \cite{DM94}).

The interpretation proposed here stems, on the other hand, only from two 
assumptions: $M/L = const.$ and $M_c \sim const.$. Both of them have natural
empirical support from the GCFP and (\lrc -- \muv) correlations respectively.
Furthermore the predicted scale law $\sigma \propto I_0^{0.25}$ give a good
match with the observed slope of the relation ($\sigma \propto
I_0^{0.32\pm0.05}$
from a sample composed solely of KM clusters). I will show below that tiny
differences between the observed and predicted slopes can be easily explained
with simple arguments well rooted in the standard theory of dynamical evolution
of globular clusters.

Since the standard interpretation of 
the $I_0 ~vs.~ \sigma$ scale law (\cite{DM94}) is
strictly coupled with that of the $L ~vs.~ \sigma$ relation, the new
explanation of the former cast some doubt on the whole framework proposed by
\cite{DM94} for these correlations. 

Finally, the proposed scenario allows a considerable economy of hypothesis,
showing that the {\em same} interpretation succesfully explains {\em three}
correlations. 

\subsection{Evolutionary effects}

It is widely accepted that dynamical evolution should in general {\em stretch}
the range of properties of a globular clusters system (\cite{DM94}, \cite{mh97} and references
therein). So, the first conclusion we can draw inspecting fig. 3 with an
``evolutionary perspective'' is that the range of $M_c$ covered by Galactic
globulars at an early time was narrower  than today's 
and - consequently - the FSL
correlations were surely more significant and stronger in 
the past\footnote{The marginal evidence that the less evolved clusters (OC)
are less dispersed around the FSL with respect to more evolved ones (IC, see
sec. 2) provide some independent support to this view.} 
This
suggests that significant primordial cosmic scatter in just {\bf one} of the
\s parameters can be at the origin of the range covered by the other two.
For instance, assuming $M/L = const.$ and $M_c \sim const.$, a range of \rc
would immediately produce corresponding ranges of \si and \muv via relations
similar to equations (\ref{eq7}) and (\ref{eq8}).

Except for episodic phenomena (as disk and bulge shocks or encounters with
giant molecular clouds; see \cite{BVFF}) the general trend of the dynamical evolution
of globular clusters can be crudely resumed as `` shrinking toward more and more
concentrated configurations till the onset of the core collapse'' (\cite{mh97},
Spitzer 1997). During core contraction, core density (and, consequently, 
surface brightness) and velocity dispersion
increase. 
Furthermore, it is firmly established that the progressive
shrinking is accompanied by continuous decrease of the core mass (\cite{cw}, 
\cite{cohn}, \cite{mh97}). 

Based on these simple prescriptions it is possible to find how GCs are expected
to move -- driven by evolution -- in the planes of fig. 3. In each of the three
panels a couple of arrows enclose (within the acute angle between them)
the possible directions of the {\em evolution vector} (EV), i.e. the possible 
lines along which clusters can move in the plane due to the general evolutionary 
trend described above. 

First of all, it should be noted that, given the likely directions of EVs in
each plane, the effect of dynamical evolution on the FSL correlations is not
expected to be dramatic. This is particulary true for the  (\lrc -- \muv)
correlation. In this plane a theoretical estimate of the EV is available: 
$\rho_0 \propto r_c^{-2.23}$ (\cite{cohn}), corresponding to 
$I_0 \propto r_c^{-1.3}$, i.e. not far from parallel to the ${\rm log} M_c =const.$
line in the considered plane (see also \cite{DM94}). 

To attempt further interpretation of possible evolutionary effects it must be
recalled that significant correlations occur between total mass (tracked by
$M_V$) and both \si and \muv (\cite{DM94}) in the sense that brighter cores and/or
higher velocity dispersion are generally associated with more massive clusters.

So, if any (marginal) trend is present {\em within} the $3 < {\rm log} M_c < 5$ 
region of each
plane of fig. 3, it appears as follows: the distribution of the points is, in
average, farther from the ${\rm log} M_c = 5$ line for cluster with lower total mass.
This would be in rough accord with theoretical expectations, since less massive
clusters are predicted to evolve faster toward core collapse due to two-body
relaxation (\cite{spitz}; see also \cite{v97}, fig. 5).      

\section{Summary and Conclusions}

The distribution of the Galactic KMCs into the \s is very similar to a straight
line lying onto the Fundamental Plane of Globular Clusters. I have shown that
{\em the simple assumptions }(supported by observations) 
{\em $M/L \simeq const.$ and
$M_c \sim const.$ provide a simultaneous explanation for all the
three mono-variate correlations present in the data} (and also for GCFP
correlations). In particular this provides
a new and more satisfactory interpretation of the correlation between \muv and
\lsi.

Some interesting by-products have also been obtained: {\em a)} the thickness of
the GCFP and of FPEG at any mass are very similar; 
{\em b)} the range of M/L covered by GCs is constrained to
be significantly less than an order of magnitude with a fully 
{\em model-independent} procedure; 
{\em c)} the current definition of observables, coupled with a strong
$M/L \simeq const.$ constraint, create a condition in which even if 
the $M_c$ range of GC would encompass 4 order of magnitude, we would still
observe some significant correlation in the planes (\lrc -- \muv) and (\muv --
\lsi).

In my opinion, the most far reaching conclusion of the present analysis is that
globulars {\em were distributed along a Fundametal Straight Line in S-space} at
early times. This conclusion provides a very useful constraint for models of
globular cluster formation, i.e. globulars were born with nearly 
{\em the same core mass} (or rapidly settled to this status), 
within a range that was probably much narrower than the one presently observed. 
Present knowledge of the formation of globular clusters is rather poor (see
\cite{fr85}, \cite{vp95}, \cite{hp94}, \cite{mh97}) and any
observational constraint on initial conditions of this system has to be
considered very valuable. 


George Djorgovski is warmly thanked for having introduced me to the mysteries
of Principal Component Analysis.
I am indebted to Paolo Montegriffo, Enrico Vesperini, Stefano Sandrelli and 
Flavio Fusi Pecci for many useful discussions.

Very special thanks are owed to Luca Ciotti for many insightful discussions 
and a critical reading of draft versions of the manuscript.

Mrs Paola Ballanti is warmly thanked for her professional expertise in english
language.   

\end{document}